\documentclass[aps,prl,reprint,superscriptaddress,preprintnumbers]{revtex4-2}

\newif\ifpublic\publictrue

\usepackage{fancyhdr}
\ifpublic\else
\fi
\newif\ifworking\workingtrue

\usepackage{mathtools,mathrsfs,amsbsy,amssymb,latexsym,amsfonts,amscd,amsmath,multirow}
\usepackage{bm}
\allowdisplaybreaks[4]
\usepackage{graphicx}
\usepackage[usenames,dvipsnames]{color}
\usepackage[normalem]{ulem}
\usepackage{natbib}
\usepackage{xcolor}
\usepackage{comment}

\definecolor{linkcolor}{rgb}{0,0,0.6}

\usepackage[pdftex,colorlinks=true,
pdfstartview=FitV,
linkcolor= linkcolor,
citecolor= linkcolor,
urlcolor= linkcolor,
hyperindex=true,
hyperfigures=false]
{hyperref}

\usepackage{tikz}
\usetikzlibrary{arrows.meta,calc,decorations.markings,patterns}

\newcommand{\dd}{\mathrm{d}}

\newcommand{\GG}{\mathrm{G}}

\newcommand{\SO}[1]{\text{SO}(#1)}
\newcommand{\Sp}[1]{\text{S}^{#1}}

\newcommand{\calR}{\mathcal{R}}

\newcommand{\vol}{\text{vol}}

\newcommand{\gYM}{g_{\text{YM}}}

\newcommand{\dual}{\text{dual}}

\begin{document}
\pagestyle{plain}

\title{Scaling similarity and generalised conformal symmetry in D2-brane holography}

\author{Andrea Conti}
\email{contiandrea@uniovi.es}
\affiliation{Departamento de F\'isica, Universidad de Oviedo, c/ Leopoldo Calvo Sotelo, 18, 33007 Oviedo, Spain}
\affiliation{Instituto Universitario de Ciencias y Tecnolog\'ias Espaciales de Asturias (ICTEA), Calle de la Independencia 13, 33004 Oviedo, Spain}

\author{Adolfo Guarino}
\email{adolfo.guarino@uniovi.es}
\affiliation{Departamento de F\'isica, Universidad de Oviedo, c/ Leopoldo Calvo Sotelo, 18, 33007 Oviedo, Spain}
\affiliation{Instituto Universitario de Ciencias y Tecnolog\'ias Espaciales de Asturias (ICTEA), Calle de la Independencia 13, 33004 Oviedo, Spain}

\author{Ricardo Stuardo}
\email{ricardostuardotroncoso@gmail.com}
\affiliation{Institute for Theoretical Physics and Leuven Gravity Institute, KU Leuven, Celestijnenlaan 200D, B-3001 Leuven, Belgium.}

\begin{abstract}

We study D2-brane holography within the purely dilatonic sector of four-dimensional maximal $\textrm{ISO}(7)$-gauged supergravity, obtained as a consistent truncation of type IIA supergravity on $\textrm{S}^{6}$. In the dual frame, the four-dimensional Lagrangian resulting from the reduction of the massless IIA theory exhibits scaling similarity, thus reflecting the generalised conformal symmetry of the dual three-dimensional maximally supersymmetric Yang–Mills theory. Using an auxiliary spacetime construction, we show that the holographic CFT$_3$ fixed points and RG-flows known in the massive IIA theory have natural counterparts in the massless IIA theory: QFT$_3$s with generalised conformal symmetry connected by generalised RG-flows, which we construct numerically. We also revisit the Coulomb branch solutions and propose an extension of Gubser's singularity criterion to supergravity theories exhibiting scaling similarity.

\end{abstract}


\maketitle

\section{Introduction}

The gauge/gravity correspondence provides a powerful framework for studying strongly-coupled gauge theories through their gravitational duals. In the case of three-dimensional (3D) maximally supersymmetric Yang--Mills (SYM) theory, precise holographic studies substantially benefit from the consistent truncation of ten-dimensional (10D) \textit{massless} type IIA supergravity on the six-sphere \cite{Hull:1988jw}, $\Sp{6}$, down to four-dimensional (4D) maximal supergravity with an electric $\textrm{ISO}(7)$ gauging \cite{Hull:1984yy}. Such a truncation ensures that every solution of the lower-dimensional theory uplifts to an exact ten-dimensional solution. The accompanying embedding formulae are equally important, as they explicitly relate lower-dimensional fields to their higher-dimensional origins. It is also well-known that type IIA supergravity admits a mass deformation, $m$, as originally found by Romans in \cite{Romans:1985tz}. The corresponding \textit{massive} type IIA supergravity still admits a consistent truncation on S$^6$ \cite{Guarino:2015vca}, yielding this time a four-dimensional maximal supergravity with a dyonic $\textrm{ISO}(7)$-gauging \cite{Guarino:2015qaa}. By setting the Romans mass parameter to zero, \textit{i.e.} $m=0$, the embedding formulae proposed in \cite{Guarino:2015vca} consistently reduce to those of the massless theory. 

In the massless case, $m=0$, one specific scalar in the $\textrm{ISO}(7)$ supergravity becomes a runaway direction of the scalar potential. Such a scalar, we denote it $\sigma$, descends from the ten-dimensional dilaton -- this becomes manifest in the (D2-brane) dual frame of \cite{Kanitscheider:2008kd} -- and its running codifies the running of the gauge coupling $\gYM^{2}$ of the dual non-conformal SYM theory which, in three dimensions, has scaling dimension one. Despite being non-conformal, the SYM theory living on the worldvolume of a stack of D2-branes in flat spacetime exhibits a so-called \textit{generalised conformal symmetry} (GCS) \cite{Jevicki:1998yr,Jevicki:1998ub}. The terminology reflects the fact that the conformal transformations are extended to act on the Yang--Mills coupling, which is treated as a spurion with appropriate scaling dimension. Although the coupling is not a dynamical degree of freedom, assigning it a transformation under the conformal group provides a useful way of keeping track of the explicit breaking of conformal symmetry. Building on this GCS, \cite{Kanitscheider:2008kd} formulated the corresponding generalised conformal \textit{structure} in terms of Weyl transformations of background fields and derived the associated Ward identities in the holographic description.

Turning on the Romans mass, $m\neq 0$, stabilises the scalar $\sigma$, thus allowing the $\textrm{ISO}(7)$ supergravity to admit AdS$_4$ solutions. The corresponding CFT$_3$ duals are strongly-coupled Chern--Simons-matter theories with gauge group $\textrm{SU}(N)$ and Chern--Simons (CS) level $k=(2\pi\ell_s) m$ determined by the Romans mass \cite{Guarino:2015jca}. The emergence of these strongly-coupled fixed points in the infrared (IR) regime of three-dimensional SYM can be understood from the presence of a relevant Chern--Simons deformation. Indeed, $\gYM^2$ grows towards the IR, so that the Yang--Mills term becomes irrelevant, leaving a Chern--Simons-matter theory at the fixed point.

In this note, we argue that the holographic picture familiar from the massive case has a natural massless counterpart. In the massive theory, CFT$_3$ fixed points and RG-flows interpolating between them are replaced, in the massless case, by QFT$_3$s with generalised conformal symmetry and generalised RG-flows connecting them. We demonstrate this correspondence in a simple sector of the $\textrm{ISO}(7)$ supergravity which retains the seven dilatons (associated with Cartan generators) of the coset space $\textrm{E}_{7(7)}/\textrm{SU}(8)$ spanned by the scalars of maximal supergravity. Importantly, this sector uplifts to massive/massless type IIA backgrounds with only a dual magnetic flux $F_{6}=\star{F}_{4}$ turned on. Whether and how these results can be extended to more general sectors, for example, including axions (which turn on a magnetic flux $H_{3}$) and vectors, remains an open question to us \footnote{Still we have verified that the results here extend straightforwardly to the STU-model of \cite{Guarino:2017pkw} with vanishing axions.}. We return to this issue in the discussion section.

Finally, we have made extensive use of the auxiliary spacetime construction introduced in \cite{Biggs:2023sqw}, see also \cite{Bobev:2025idz,BBGM}, to motivate and formulate our proposals. This construction also motivates us to propose a criterion for acceptable singularities in supergravity solutions holographically describing (deformations of) QFT$_3$s with generalised scaling symmetry. It can be viewed as a direct extension of Gubser's criterion \cite{Gubser:2000nd} to gravity theories holographically dual to QFTs with a generalised conformal symmetry. The note closes with several complementary appendices.

\section{A simple model}\label{sec:seven-dilaton_model}

As stated in the introduction, we focus on the purely dilatonic sector of the maximal ISO(7) gauged supergravity of \cite{Guarino:2015qaa}. This can be obtained from the $\mathbb{Z}_{2}^{3}$-invariant sector put forward in \cite{Guarino:2019snw} upon setting to zero the (seven) axions, thus resulting in what we call the seven-dilaton model. The model is just an Einstein-scalar model involving $1+6$ scalars $(\sigma,\lambda_{a})$ with $a=1, \ldots,6$. However, it proves convenient (see \textit{e.g.} \cite{Cvetic:2000zu}) to express the Lagrangian in terms of $\sigma$ and seven auxiliary scalars $X_{I}$, $I=1,\ldots,7$, which satisfy the constraint $\prod_{I} X_{I} =1$. The latter are defined in terms of the physical scalars $\lambda_{a}$ using the weights of the fundamental representation of SL(7), $(b_{I})^{a}$, as $X_{I} \equiv e^{-\frac{1}{2} (b_{I})^{a} \lambda_{a}}$ (see Appendix~\ref{app:scalar_parameterisation}). Then, the seven-dilaton model is described by a Lagrangian of the form
\begin{equation}
\label{eq:7DilatonLagrangian}
\begin{array}{rcl}
\mathcal{L}_{4} &=& \left(\dfrac{R_{4}}{2} - V \right)\star_{4}1 \\
&-& \dfrac{1}{4}\dd\sigma\wedge\star_{4}\dd \sigma -\dfrac{1}{8} \displaystyle\sum_{I} X_{I}^{-2} \dd X_{I} \wedge \star_{4} \dd X_{I} \ ,
\end{array}
\end{equation}
where the scalar potential reads
\begin{equation}
\label{eq:Potential7Dilaton}
V = \frac{g^{2}}{8} \, e^{\frac{\sigma}{\sqrt{7}}} \, V_X + \frac{m^{2}}{8} \, e^{\sqrt{7}\sigma} \ ,
\end{equation}
and contains an $X$-dependent piece
\begin{equation}
\label{eq:V_lambda}
V_{X} = 2\sum_{I} X^{2}_{I} - \left(\sum_{I} X_{I}\right)^{2} \ .
\end{equation}
In the massless case, $m=0$, the scalar potential in (\ref{eq:Potential7Dilaton}) takes the factorised form $V = \frac{g^{2}}{8} \, e^{\frac{\sigma}{\sqrt{7}}} \, V_X$. In this case, the splitting $(\sigma,X_{I})$ of the scalars becomes convenient to carry out holographic computations in the dual frame of \cite{Kanitscheider:2008kd}: It isolates the scalar coming from the ten-dimensional dilaton, $\sigma$, from the scalars $X_{I}$ corresponding to squashing of the six-sphere.

The function $V_{X}$ in \eqref{eq:V_lambda} has $1+7$ extrema by itself. There is a single extremum that preserves the SO(7) isometry group of the round six-sphere. The location of this extremum and the associated value of $V_{X} $ are given by
\begin{equation}
\label{eq:XSO7vacuum}
X^{(0)}_{I} = 1 
\quad, \quad
V^{(0)}_{X} = -35 \ .
\end{equation}
There are seven additional degenerate extrema preserving a subgroup $\textrm{SO}(6)\subset \textrm{SO}(7)$ of the isometry group of the six-sphere. These are labeled by $\mathfrak{I}=1,\ldots,7$, and have
\begin{equation}
\label{eq:XSO6vacuum} 
X^{(0)}_{I} = \begin{cases}
         &2^{\frac{12}{7}} \quad \text{ for $I=\mathfrak{I}$}  \\
         &2^{-\frac{2}{7}} \quad \text{for $I \neq \mathfrak{I}$}  \\
        \end{cases}
        \quad,\quad
V^{(0)}_{X} = -2^{\frac{17}{7}}\, 7  \ .
\end{equation}
We will use the above set of extrema of $V_{X}$ as a seed to construct various types of solutions in the theory (\ref{eq:7DilatonLagrangian}).

\section{Generalised conformal solutions}\label{sec:genconfsol}

The constant scalar values in (\ref{eq:XSO7vacuum}) and (\ref{eq:XSO6vacuum}) can be used to construct both conformal and generalised conformal solutions.

\subsection{The conformal case}

Before presenting the generalised conformal solutions, let us review the ordinary conformal solutions of (\ref{eq:7DilatonLagrangian}). Conformality requires an AdS$_{4}$ spacetime, \textit{i.e.} $\dd s_{4}^2 = \dd s^{2}_{\textrm{AdS}_{4}}$, and the scalars to be fixed at an extremum of the scalar potential (\ref{eq:Potential7Dilaton}). These solutions are only possible in the massive case $m \neq 0$ since, in the massless case, the scalar $\sigma$ becomes a runaway direction of the potential (\ref{eq:Potential7Dilaton}). Due to the form of the potential in \eqref{eq:Potential7Dilaton}, finding the extrema of $V$ reduces to finding those of $V_{X}$, as the value of the scalar $\sigma$ is simply fixed to
\begin{equation}
\label{eq:sigma_VEV}
\sigma = \frac{\sqrt{7}}{6}\log\left(-\frac{g^{2}}{7m^{2}} \, V^{(0)}_{X}  \right) \ ,
\end{equation}
with $V^{(0)}_{X}<0$. Plugging (\ref{eq:sigma_VEV}) back into (\ref{eq:Potential7Dilaton}) yields
\begin{equation}
V^{(0)} =  -\frac{3}{7^{7/6} 4} \left(\frac{g^{7}}{m}\right)^{\frac{1}{3}} \left(-V_{X}^{(0)} \right)^{\frac{7}{6}} < 0 \ ,
\end{equation}
which then sets the radius $\ell^{2}=-3/V^{(0)}$ of the AdS$_{4}$ spacetime metric. While the simple model (\ref{eq:7DilatonLagrangian}) admits only the single SO(7)-symmetric and the seven SO(6)-symmetric AdS$_{4}$ solutions specified by (\ref{eq:XSO7vacuum}) and (\ref{eq:XSO6vacuum}), both of them being non-supersymmetric, many additional AdS$_{4}$ solutions (with and without supersymmetry) have been found in extended setups that also include axions \cite{Guarino:2015qaa,Guarino:2019jef,Guarino:2019snw,Bobev:2020qev}. The uplift of the SO(7)- and SO(6)-symmetric AdS$_{4}$ solutions to ten-dimensional massive IIA supergravity is direct using (\ref{eq:Uplift7Dilaton}).

\subsection{The generalised conformal case}
\label{sec:flatDW}

In the massless case $m=0$ the potential (\ref{eq:Potential7Dilaton}) does not allow for AdS$_{4}$ solutions with constant $\sigma$, as it has a running potential \footnote{Still, massless IIA solutions with a running dilaton and an AdS$_{4}$ factor in string frame have been discussed in \cite{Gaillard:2010gy}.}. As discussed in the introduction, this is to be expected as the running of $\sigma$ realises the running of the gauge coupling $\gYM^2$ of the dual non-conformal SYM theory on $\mathbb{R}^{1,2}$. Still, it allows for a class of generalised conformal solutions which we describe now.

In order to find gravity duals, we consider \textit{flat-sliced} domain-wall solutions with constant $X_{I}=X^{(0)}_{I}$. The solutions are non-conformal but still adopt a generalised conformal form
\begin{equation}
\label{eq:D2braneConstantLambda}
\dd s^{2}_{4} = \calR^{2}\left(\frac{\sqrt{2}}{g}\right)^{2} e^{-\frac{\sigma}{\sqrt{7}}} \, \dd s^{2}_{\textrm{AdS}_{4}}
\,\, , \,\,
e^{-\frac{\sigma}{\sqrt{7}}} = \left( \calR \frac{\sqrt{2}}{g} \, \ell\,  r\right)^{z} \ ,
\end{equation}
with $z=1/3$, and
\begin{equation}
\label{eq:AdS4_metric}
\dd s^{2}_{\textrm{AdS}_4} = \ell^{2} \left(r^{2} \dd x^{2}_{1,2} + \frac{\dd r^{2}}{r^{2}}\right)  \ ,
\end{equation}
where $ \dd x^{2}_{1,2} $ is the line element of $\mathbb{R}^{1,2}$. We have introduced the constant $\calR = 2/(5-p)$ with $p=2$, and the AdS$_{4}$ line element in (\ref{eq:AdS4_metric}) with radius 
\begin{equation}
\label{l2_generalised_conformal}
\ell^{2} = -\frac{35}{V^{(0)}_{X}}  \ .
\end{equation}
While the SO(7)-symmetric generalised conformal solution associated with the scalar values in (\ref{eq:XSO7vacuum}) is $1/2$-BPS supersymmetric, the SO(6)-symmetric ones associated with (\ref{eq:XSO6vacuum}) are non-supersymmetric. When uplifted to massless IIA using (\ref{eq:uplift_generalised_conformal}), the SO(7)-symmetric configuration recovers the near-horizon limit of a stack of D2-branes in flat spacetime. This solution is $1/2$-BPS supersymmetric and the SO(7) symmetry accounts for the R-symmetry group of $\mathcal{N}=8$ SYM theory. The SO(6)-symmetric configurations describe a non-supersymmetric mass deformation of the $\mathcal{N}=8$ SYM theory.

Let us point out that the massless solutions above admit a continuous generalisation to the massive case, $m\neq 0$, by replacing the flat-sliced domain-wall Ansatz with an AdS$_3$-sliced one, such that (\ref{eq:D2braneConstantLambda})-(\ref{eq:AdS4_metric}) are recovered smoothly in the $m \to 0$ limit. These solutions are presented in Appendix~\ref{app:curved_DW_mIIA}. While the uplift of the SO(7)-symmetric configuration gives the $1/2$-BPS supersymmetric D2-O8 system studied in \cite{Dibitetto:2018ftj,Legramandi:2020txf},  the uplift of the SO(6)-symmetric configurations gives a non-supersymmetric deformation of that system.

\section{On the generalised conformal symmetry}
\label{sec:generalised_conformal}

In the massless case, $m=0$, the Lagrangian in (\ref{eq:7DilatonLagrangian}) can be moved to a four-dimensional version (see \cite{Boonstra:1998mp}) of the \textit{dual frame} introduced in \cite{Kanitscheider:2008kd}, sometimes also referred to as \textit{brane frame}. This is achieved by introducing a four-dimensional dual metric as
\begin{equation}
\label{eq:4D_metric_BraneFrame}
\dd s_{4}^{2} = \calR^{2} \left(\frac{\sqrt{2}}{g}\right)^{2} e^{-\frac{\sigma}{\sqrt{7}}}\, \dd s_{4,\textrm{dual}}^{2}\ , 
\end{equation}
and also performing the field redefinition 
\begin{equation}
\label{hat_sigma_def}
\sigma = \frac{2\sqrt{7}}{5}\, \hat{\sigma} \ .
\end{equation}
After this, the Lagrangian in (\ref{eq:7DilatonLagrangian}) takes the dual form
\begin{equation}
\label{eq:4D_Lagrangian_dual}
\begin{aligned}
\mathcal{L}_{4,\text{dual}} & = \left( \frac{\calR}{g}\right)^{2} e^{\gamma \hat{\sigma}} \left[ \left(R_{4,\textrm{dual}} - \frac{\calR^{2}}{2} V_{X}\right) \hat{\star}_{4} 1 \right. \\
& \left. + \, \beta \, \dd \hat{\sigma} \wedge \hat{\star}_{4} \dd \hat{\sigma} - \dfrac{1}{4} \displaystyle\sum_{I} X_{I}^{-2} \dd X_{I} \wedge \hat{\star}_{4} \dd X_{I}  \right] ,
\end{aligned}
\end{equation}
with $\calR = \frac{2}{3}$, $\gamma = -\frac{2}{5}$ and $\beta = -\frac{8}{25}$, and where the $\hat{\star}_{4}$ operation is defined with respect to the dual metric. The fact that the dual Lagrangian (\ref{eq:4D_Lagrangian_dual}) factorises neatly, with an overall factor of $e^{\gamma \hat{\sigma}}$, follows from the factorised form of the scalar potential in \eqref{eq:Potential7Dilaton} when $m=0$. Therefore, we will refer to $V_{X}$ in (\ref{eq:4D_Lagrangian_dual}) as the \textit{brane frame scalar potential}.

The dual Lagrangian (\ref{eq:4D_Lagrangian_dual}) can be viewed as an extension of the single-field model in \cite{Boonstra:1998mp,Kanitscheider:2008kd} to include the seven dilatons of the maximal ISO(7) supergravity. Under $\hat{\sigma} \rightarrow \hat{\sigma} + c$, with constant $c$, (\ref{eq:4D_Lagrangian_dual}) undergoes a \textit{scaling similarity} transformation $\mathcal{L}_{4,\text{dual}} \rightarrow e^{\gamma c}\mathcal{L}_{4,\text{dual}}$. Even though this is not a symmetry of the Lagrangian, it is a symmetry of the dual frame equations of motion. Such a structure of the four-dimensional theory is a consequence of the fact that it describes a consistent truncation of massless type IIA on the six-sphere around the $1/2$-BPS and SO(7)-symmetric D2-brane background. This is also true for supergravity theories obtained as consistent truncations on ($8-p$)-spheres around D$p$-brane backgrounds, for $p\leq4$ and $p\neq 3$: in these cases it is also possible to move to the dual frame of \cite{Kanitscheider:2008kd}, where the theory manifests scaling similarity. As the worldvolume theory of D$p$-branes ($p\leq4$ and $p\neq 3$) preserves a generalised conformal symmetry, it is natural to expect that solutions of supergravities with scaling similarity holographically describe (deformations of) QFTs preserving a generalised conformal symmetry.

\subsection{Auxiliary \texorpdfstring{AdS$_{4+\eta}$}{AdS 4+eta} gravity and fixed points}

The theories with scaling similarity discussed above can alternatively be obtained by dimensional reduction of a ($p+2+\eta$)-dimensional auxiliary theory with similarity exponent $\eta = \frac{(p-3)^{2}}{5-p}$ ($p\leq4$ and $p\neq 3$) \cite{Biggs:2023sqw}. In what follows, we specialise to the D2-brane case for which $p=2$ and $\eta=1/3$. The uplift proceeds with a Kaluza--Klein (KK)  Ansatz for the auxiliary metric
\begin{align}
\label{eq:aux_metric}
\dd s^{2}_{4+\eta} &= g^{(\dual)}_{\mu\nu} \dd x^{\mu} \dd x^{\nu} + e^{-\frac{4}{5\eta}\hat{\sigma}} \delta_{ij} \dd \xi^{i} \dd \xi^{j} \ ,
\end{align}
where $\xi^{i}$ are coordinates on $\mathbb{R}^{\eta}$, and gives an action for the auxiliary theory of the form
\begin{align}
\label{eq:auxAdSGrav}
S^{(\text{aux})} &= \frac{\calR^{2}}{g^{2}} \vol_{\eta}^{-1} \int \dd^{4}x\,  \dd^{\eta} \xi \sqrt{-g_{4+\eta}} \\
& \left( R - \frac{1}{4} \sum_{I} X_{I}^{-2} (\partial X_{I})^2 - \frac{\calR^{2}}{2} V_{X} \right) \, , \nonumber
\end{align}
with $\vol_{\eta}$ being the volume of the extra $\eta$-dimensions. In the auxiliary ($4+\eta$)-dimensional theory there is no $\hat{\sigma}$ scalar and  $V_{X}$ plays the role of the scalar potential. Notably, this potential can be obtained from a real superpotential $W_{\lambda} \equiv \sum_{I} X_{I}$ with $X_{I} \equiv e^{-\frac{1}{2} (b_{I})^{a} \lambda_{a}}$ as
\begin{equation}
\label{eq:VfromW}
V_{X} = \sum_{a} \left( \frac{\partial W_{\lambda}}{\partial \lambda_{a}} \right)^{2} - \frac{D-1}{2(D-2)} W^{2}_{\lambda}  \ ,
\end{equation}
with $D=4+\eta$ and $\eta=1/3$.

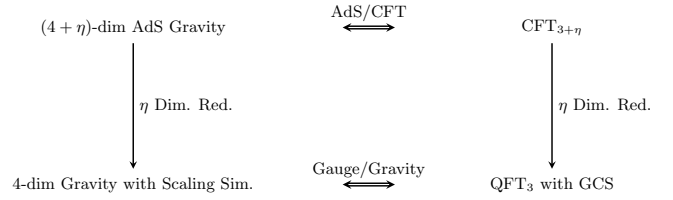
\begin{figure}[t]
    \centering
    \resizebox{0.48\textwidth}{!}{%
    \begin{tikzpicture}
        \node (auxAdS) at (-4,1) (TextNode) {$(4+\eta)$-dim AdS Gravity};
        \node (auxCFT) at (4,1) (TextNode) {CFT$_{3+\eta}$};
        
        \node (scaleCov) at (-4,-2) (TextNode) {$4$-dim Gravity with Scaling Sim.};
        \node (genConf) at (4,-2) (TextNode) {QFT$_{3}$ with GCS};
        
        \draw [thick,-stealth] (-4,0.7) -- (-4,-1.7) 
            node [midway,right] (TextNode) {$\eta$ Dim. Red.};
        \draw [thick,-stealth] (4,0.7) -- (4,-1.7) 
            node [midway,right] (TextNode) {$\eta$ Dim. Red.};
        \draw[thick,stealth-stealth,double] (0,1) -- (1,1) 
            node [midway,above] (TextNode) {AdS/CFT};
        \draw[thick,stealth-stealth,double] (0,-2) -- (1,-2) 
            node [midway,above] (TextNode) {Gauge/Gravity};
    \end{tikzpicture}%
    }
\caption{Relation between the auxiliary AdS$_{4+\eta}$/CFT$_{3+\eta}$ duality and the gravity with scaling sim./QFT$_3$ correspondence, where the QFT$_3$ exhibits a generalised conformal symmetry. The latter correspondence is obtained by dimensional reduction along the $\eta$ directions.}
\label{fig:AuxAdSPicture}
\end{figure}

At this level, one can analyse different solutions of the model, \textit{e.g.} (asymptotically) AdS$_{4+\eta}$ solutions, as holographic duals to (deformations of) an auxiliary CFT$_{3+\eta}$ \cite{Biggs:2023sqw,Bobev:2025idz}. Then, by dimensionally reducing on the $\eta$ directions, one recovers the holographic duality between gravity theories with scaling similarity and QFTs with a generalised conformal symmetry. From this perspective, the generalised conformal symmetry of the QFT$_3$ descends from the conformal symmetry of the auxiliary CFT$_{3+\eta}$ through dimensional reduction. Or, in other words, the generalised conformal symmetry inherits its set of ``fixed points" from the set of AdS$_{4+\eta}$ solutions of the auxiliary theory. The picture is summarised in Figure \ref{fig:AuxAdSPicture}.

The above invites to speculate -- still we do not claim any physical meaning of the AdS$_{4+\eta}$ auxiliary solutions -- on the stability of the generalised conformal solutions of Section~\ref{sec:flatDW} in terms of the normalised mass spectra of the AdS$_{4+\eta}$ solutions of the auxiliary theory (\ref{eq:auxAdSGrav}).  
Such AdS$_{4+\eta}$ solutions are obtained upon extremisation of the brane frame scalar potential $V_{X}$ and have radius $\ell^{2}_{(\textrm{aux})}$. Notably, the AdS$_{4+\eta}$ solutions are the uplift of the generalised conformal solutions (\ref{eq:D2braneConstantLambda})-(\ref{eq:AdS4_metric}) to the auxiliary theory. Then
\begin{equation}
\ell^{2}_{(\textrm{aux})} = \ell^{2} = -\frac{35}{V^{(0)}_{X}} \ .
\end{equation}
For the SO(7)-symmetric solution describing the D2-brane, the normalised masses (and multiplicities) are $m^{2} \ell^{2}_{(\textrm{aux})}=-\frac{8}{3} \,\, (\times 6)$. Note that all the masses lie above the Breitenlohner--Freedman (BF) bound, $m^{2} \ell^{2}_{(\textrm{aux})} \ge -\frac{25}{9}$, for perturbative stability in AdS$_{4+\eta}$ with $\eta=1/3$ \cite{Breitenlohner:1982jf}. This is consistent with the SO(7)-symmetric solution being supersymmetric. On the contrary, for the non-supersymmetric and SO(6)-symmetric solutions, the normalised masses read $m^{2} \ell^{2}_{(\textrm{aux})}=-\frac{10}{3} \,\, (\times 5) \,\, , \,\, \frac{20}{3} \,\, (\times 1)$. The value $-\frac{10}{3}$ violates the BF bound potentially signalling a pathology of the corresponding generalised conformal solutions.

\subsection{Generalised fixed points and RG-flows} \label{sec:genRGflow}

\begin{figure}[t]
        \centering
        \includegraphics[width=0.95\linewidth]{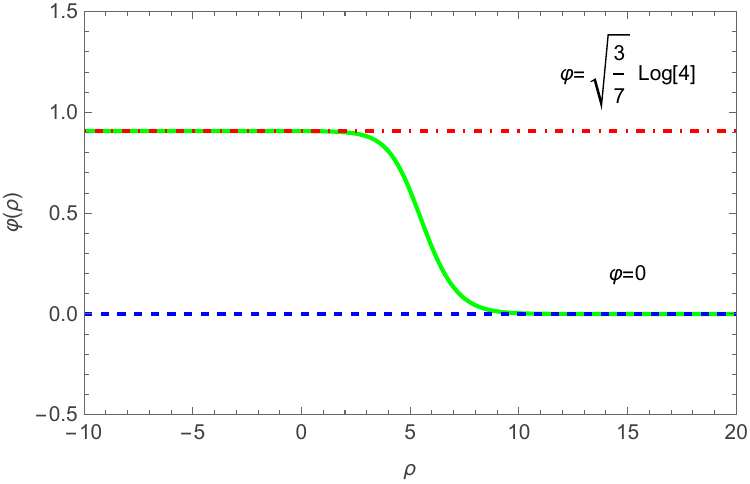}
        \includegraphics[width=0.95\linewidth]{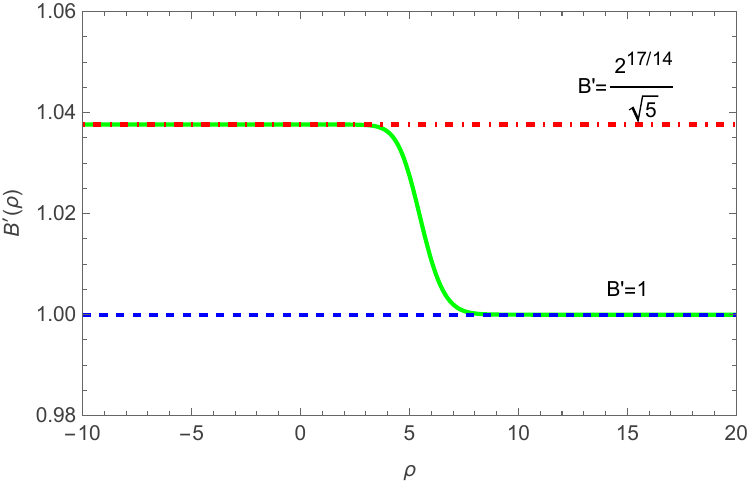}
        \caption{Top: Running of the SO(6)-invariant scalar $\varphi$. \qquad Bottom: Radial derivative of the metric fuction $B(\rho)$, the asymptotic value being $1/\ell$. Dashed/Blue line = UV fix point; Dot-dashed/Red line = IR fixed point, Green/Solid line = RG-flow.}
        \label{fig:placeholder}
\end{figure}

We can push the discussion a bit further and consider not only a correspondence between AdS$_{4+\eta}$ solutions of the auxiliary theory and four-dimensional generalised conformal solutions, but also to expect the existence of generalised RG-flows connecting generalised conformal solutions. In particular, for the (supersymmetric) SO(7)-symmetric extremum of $V_{X}$ in (\ref{eq:XSO7vacuum}) and the (non-supersymmetric) SO(6)-symmetric extrema in (\ref{eq:XSO6vacuum}), we observe that
\begin{equation}\label{eq:SO7andSO6radii}
\ell^{2}_{\text{SO(7)}} > \, \ell^{2}_{\text{SO(6)}} \ ,
\end{equation}
which a priori allows for non-supersymmetric generalised RG-flows between the QFT$_{3}$ duals exhibiting generalised conformal symmetry. We have constructed such RG-flows numerically, see Figure~\ref{fig:placeholder}, using an Ansatz of the form
\begin{equation}
\label{eq:DW-ansatz_main_text}
\dd s_{4}^{2} = \calR^{2}\left(\frac{\sqrt{2}}{g}\right)^{2} e^{-\frac{\sigma}{\sqrt{7}}}
\left(e^{2 B(\rho)} \dd x_{1,2}^{2} + \dd \rho^{2}\right) \ ,
\end{equation}
with $\varphi(\rho)$ being the SO(6)-invariant scalar (see eq.(\ref{eq:XtoSectors}) below with $n=6$). Moreover, the scalar $\sigma$ relates to the metric function $B$ as $\sigma = -(\sqrt{7}/3) B + \sigma_{0}$, with $\sigma_{0}$ being a real integration constant, so that the generalised 4D domain-walls connecting the generalised conformal solutions of Section~\ref{sec:flatDW} are in one-to-one correspondence with the domain-walls connecting AdS$_{4+\eta}$ solutions in the auxiliary theory (\ref{eq:auxAdSGrav}).

\section{Coulomb branch revisited}
\label{sec:Coulomb_branch}

For $m=0$, the general flat-sliced supersymmetric domain-wall solutions describing the Coulomb branch of three-dimensional SYM theory in flat spacetime were constructed in \cite{Cvetic:2000zu} in the context of the domain-wall/QFT correspondence \cite{Boonstra:1998mp}. In the conventions of this note, the relevant solutions of the BPS-equations (see Appendix~\ref{app:BPS_equations_7_dilaton}) are
\begin{align}\label{eq:Cvetic7Dilaton}
        \dd s^{2}_{4} & = {\cal{R}}^2 \left(\frac{\sqrt{2}}{g}\right)^2 e^{-\frac{\sigma}{\sqrt{7}}} \, H^{\frac{3}{14}}\left( \dd x^2_{1,2} + \frac{r^{\frac{2}{3}}\dd r^2}{\sqrt{H}}\right) \ , \nonumber \\
        e^{-\frac{\sigma}{\sqrt{7}}}&= \left({\cal{R}}\frac{\sqrt{2}}{g} \right)^{\frac{1}{3}} H^{\frac{1}{28}} \quad,\quad
        X_{I} = \frac{H^{1/7}}{h_I} \ ,
\end{align}
with $H = \prod_I h_I$ and $h_I = r^{\frac{4}{3}} + q_{I}^{2}$. Again $\calR = 2/3$, and the solutions involve a set of arbitrary real constants $q_{I}$. The uplift of (\ref{eq:Cvetic7Dilaton}) to massless IIA is direct using Appendix~\ref{app:IIA_uplift_7_dilatons}, and the D2-brane distribution giving rise to the geometry was given in eq.$(36)$ of \cite{Cvetic:2000zu}.

\subsection{ \texorpdfstring{$\textrm{SO}(n) \times \textrm{SO}(7-n)$}{SO(n) x SO(7-n)} symmetry enhancement}
\label{sec:Coulomb_branch_SO(n)xSO(7-n)}

Setting $q_{I} = q$, for $I=1,...,n$, and $q_{I} = 0$, for $I=(n+1),...,7$, in the general solution of (\ref{eq:Cvetic7Dilaton}), produces a symmetry enhancement to $\textrm{SO}(n) \times \textrm{SO}(7-n)$ with $n=1,...,6$. This in turn implies an identification between the scalars $X_{I}$ of the form
\begin{equation}
\label{eq:XtoSectors}
        X_{I} = 
        \begin{cases}
           & e^{-\frac{7-n}{c_{n}}\varphi} \quad \text{for $I=1,...,n$} \\
           & e^{\frac{n}{c_{n}}\varphi} \,\qquad  \text{for $I=n+1,...,7$} 
        \end{cases} \ ,
\end{equation}
where $c^{2}_{n}= \frac{7}{2} \, n \, (7-n)$. The resulting solutions only involve the scalar $\sigma$, which preserves the full $\SO{7}$ isometry group of the round six-sphere, and one scalar, we denote it by $\varphi$, responsible for the breaking of the six-sphere isometries down to $\textrm{SO}(n) \times \textrm{SO}(7-n) \subset \textrm{SO}(7)$ with $n=1,...,6$. More concretely  
\footnote{
Using (\ref{eq:XtoSectors}), the solutions (\ref{eq:Ansatz4DCoulomb})-(\ref{eq:Solution4Dv1}) solve the BPS-equations in Appendix~\ref{app:BPS_equations_7_dilaton} with a factorised real superpotential $W = \frac{g}{4 \sqrt{2}} e^{\frac{\sigma }{2 \sqrt{7}}} \, W_{\varphi}$ with $W_{\varphi} =  n \, e^{- \frac{7-n}{c_{n}} \, \varphi} + (7-n) \, e^{\frac{n}{c_{n}} \, \varphi}$. We note that $c_{n}$ is symmetric under the exchange $n \leftrightarrow (7-n)$, so the four-dimensional theory is invariant under the simultaneous transformations $\varphi \rightarrow - \varphi$ and $n \leftrightarrow (7-n)$. However, the solutions (\ref{eq:Ansatz4DCoulomb})-(\ref{eq:Solution4Dv1}) are not.
},
\begin{equation}
\label{eq:Ansatz4DCoulomb}
\dd s^{2}_{4}  =\calR^{2} \left(\frac{\sqrt{2}}{g}\right)^{2}  e^{- \frac{\sigma}{\sqrt{7}} } e^{ \frac{3 n}{2 c_n} \varphi } \left( r^2 \dd x_{1,2}^2 + \frac{\dd r^2}{r^2 \, e^{ \frac{7n}{2 c_n} \varphi } } \right) \ ,
\end{equation}
and
\begin{equation}
\label{eq:Solution4Dv1}
e^{-\frac{\sigma}{\sqrt{7}}} =  \left( \calR \frac{\sqrt{2}}{g} \right)^{\frac{1}{3}} \left(\frac{h^{n}}{r^{\frac{4}{3}(n-7)}}\right)^{\frac{1}{28}}
\,\, , \,\,
e^{7 \varphi} = \left(\frac{h}{r^{\frac{4}{3}}} \right)^{c_n} \ ,
\end{equation}
with $h = r^{\frac{4}{3}} + q^{2}$. We note in passing that this is a solution for \textit{any} value $0< n< 7$, not just $n=1,...,6$, although a type IIA uplift only exists for the integer cases. The uplift of (\ref{eq:Ansatz4DCoulomb})-(\ref{eq:Solution4Dv1}) to massless type IIA supergravity can be obtained from the formulae in Appendix~\ref{app:Coulomb_branch_SO(n)xSO(7-n)_uplift}. The resulting ten-dimensional solutions are the D2-brane version of the continuous D3-brane distributions studied in \cite{Freedman:1999gk}. As in those cases, the D2-brane distributions can be obtained from the general case in \cite{Cvetic:2000zu} by systematically taking the limit $q_{I}\to 0$ for $I=(n+1),...,7$ in the generic D2-brane distribution.

\subsection{A singularity criterion for gravity theories with scaling similarity}

Although the $\textrm{SO}(n) \times \textrm{SO}(7-n)$ Coulomb branch solutions are all singular at $r=0$ (so they are in the dual frame), when uplifted to 10D, the cases $n=1,...,5$ satisfy the Maldacena-Nunez criterion for allowed singularities \cite{Maldacena:2000mw}: the ($-g^{(E)}_{tt}$) component of the ten-dimensional Einstein frame metric does not increase as one approaches the singularity. The fact that the $n=6$ case is not an allowed singularity is consistent with the fact that the brane distribution giving rise to this configuration is not positive everywhere \cite{Cvetic:2000zu}. This exact phenomenon was encountered for D3-brane distributions preserving SO($n$) $\times$ SO($6-n$) $\subset$ SO($6$) ($n=1,...,5$) \cite{Freedman:1999gk}. In that case, the brane distribution preserving SO($5$) $\subset$ SO($6$) is also not positively defined everywhere, indicating the presence of ``ghost" D3-branes of negative charge and tension \cite{Gubser:2000nd}, so that solution is regarded as unphysical. Moreover, the five-dimensional gauged supergravity solution describing this unphysical ten-dimensional solution does not satisfy Gubser's criterion for allowed singularities \cite{Gubser:2000nd}, while the other SO($n$) $\times$ SO($6-n$) $\subset$ SO($6$) ($n=1,...,4$) solutions do. This criterion is applied at the level of gauged supergravities, and it states that large curvatures in asymptotically AdS$_{5}$ domain-walls of the form
\begin{equation}
\label{eq:domain-wall_5D}
\dd s^{2}_{5} = e^{2A(r)}\dd x^{2}_{1,3} + \dd r^{2} \ , \quad \vec{\varphi} = \vec{\varphi}(r) \ ,
\end{equation}
with $\dd x^{2}_{1,3}$ the line element of $\mathbb{R}^{1,3}$, are allowed if the scalar potential is bounded from above when evaluated on the solution. 

Going back to the D2-brane Coulomb branch solutions preserving SO($n$) $\times$ SO($7-n$) $\subset$ SO($7$), the similarities of our construction of domain-walls -- which asymptote to a theory that preserves a generalised conformal symmetry -- together with the fact that we have argued that in the dual frame $V_{X}$ should be regarded as the scalar potential, make it natural to propose an extension of Gubser's criterion to solutions of supergravity theories with scaling similarity dual to non-conformal field theories that enjoy a generalised conformal symmetry:
\begin{center}
\textit{``Large curvature singularities in domain-wall solutions of gravity theories with scaling similarity of the form (\ref{eq:4D_Lagrangian_dual}) are allowed, if the brane frame scalar potential is bounded from above in the solutions."}
\end{center}
The criterion above can be viewed as the dimensional reduction of Gubser's one applied to the auxiliary theory. 

The Coulomb branch solutions in \eqref{eq:Ansatz4DCoulomb}-\eqref{eq:Solution4Dv1} can be recast into (the four-dimensional analogue of) the form \eqref{eq:domain-wall_5D} upon an appropriate redefinition of the radial coordinate. When uplifted to the $(4+\eta)$-dimensional auxiliary theory they become asymptotically AdS$_{4+\eta}$, and applying Gubser's criterion in the auxiliary theory becomes justified. For the solutions discussed here, the $n=1,...,5$ meet this criterion, whereas the $n=6$ case does not, which is consistent with the fact that it does not satisfy the Maldacena-Nunez criteria, and also that it should be regarded non-physical due to the brane distribution being non everywhere positive.

\section{Discussion}

Within the framework of non-conformal D$p$-brane holography \cite{Boonstra:1998mp,Kanitscheider:2008kd,Biggs:2023sqw,Bobev:2025idz}, we have investigated to what extent structures familiar from CFT holography can have natural avatars in a non-conformal setting. We have investigated these questions in the specific context of D2-brane holography, focusing on the purely dilatonic sector of the $\textrm{ISO}(7)$ maximal supergravity obtained by reducing type IIA supergravity on S$^6$. In particular, we have shown that the CFT$_3$ fixed points and RG-flows of the massive IIA theory have counterparts in the massless IIA theory: these are QFT$_{3}$s with generalised conformal symmetry and generalised RG-flows connecting them.

A natural question is whether there exist supersymmetric generalised conformal solutions of the form (\ref{eq:D2braneConstantLambda})-(\ref{eq:AdS4_metric}) beyond the $\textrm{SO}(7)$-symmetric D2-brane solution. Addressing this question requires considering more general sectors of the $\textrm{ISO}(7)$ supergravity that include axions. It is therefore natural to ask whether, and in what way, axions can be incorporated into the generalised conformal symmetry. To explore this we will consider the minimal model describing the G$_{2}$-invariant sector of the $\textrm{ISO}(7)$ supergravity \cite{Guarino:2015qaa}. This sector contains the scalar $\sigma$ and its axionic partner $\chi$, and can again be obtained from the $\mathbb{Z}_{2}^{3}$-invariant sector put forward in \cite{Guarino:2019snw} upon identifying the seven chiral multiplets therein as $z_{I} = e^{-\frac{\sigma}{\sqrt{7}}}\left( \frac{\chi}{\sqrt{7}} + i\right)$. The G$_{2}$-invariant model has a Lagrangian
\begin{equation}
\label{eq:Lagrangian_G2}
\begin{aligned}
\mathcal{L}_{4} & = \left( \frac{R_{4}}{2} - V_{\GG_{2}} \right) \star_{4}1 - \frac{1}{4}\dd \sigma \wedge \star_{4} \dd \sigma \\
& - \frac{1}{4}\left(\dd \chi - \frac{\chi}{\sqrt{7}}\dd\sigma\right) \wedge \star_{4}\left(\dd \chi - \frac{\chi}{\sqrt{7}}\dd\sigma\right) \ ,
\end{aligned}
\end{equation}
with a scalar potential of the form
\begin{equation}
\begin{aligned}
V_{\GG_{2}} = \frac{g^2}{8} \, e^{\frac{\sigma}{\sqrt{7}}} V_{\chi}  + \frac{m^2}{8} e^{\sqrt{7}\sigma}  + \frac{g\, m}{4\sqrt{7}}  e^{\frac{4\sigma}{\sqrt{7}}} \chi^{3} \ ,
\end{aligned}
\end{equation}
and an axion-dependent piece
\begin{equation}
\label{V_chi}
V_{\chi} = \left( 7 \chi^{2} - 35 \right)\left( 1 + \frac{\chi^{2}}{7} \right)^{2} \ .
\end{equation}

Setting $m=0$ again factorises the scalar potential as $V_{\GG_{2}} = \frac{g^2}{8} \, e^{\frac{\sigma}{\sqrt{7}}} V_{\chi}$ with the $e^{\frac{\sigma}{\sqrt{7}}}$ factor needed to obtain a theory with scaling similarity when moving to the dual frame of Section~\ref{sec:generalised_conformal}. As perhaps expected by now, there is a non-supersymmetric and G$_{2}$-symmetric (see Appendix~\ref{app:BPS_equations_G2-sector}) generalised conformal solution with constant $ \chi =\chi^{(0)}$ and
\begin{equation}
\label{eq:confAdSAxion}
        \dd s^{2}_{4} = \calR^{2}\left(\frac{\sqrt{2}}{g}\right)^{2} e^{-\frac{\sigma}{\sqrt{7}}} \dd s^2_{\text{AdS}_4} 
        \quad,\quad
        e^{-\frac{\sigma}{\sqrt{7}}}  =  
        (\ell \, r)^{z} \ ,
\end{equation}
with $z=\frac{\sqrt{65}-7}{4}$, and where $\dd s^2_{\text{AdS}_4}$ is taken in the flat-slicing of \eqref{eq:AdS4_metric}. Due to the $(\sigma,\chi)$ mixing in the kinetic terms (\ref{eq:Lagrangian_G2}), the equations of motion no longer force $\chi^{(0)}$ to be an extremum of $V_{\chi}$. Still, a solution exists with
\begin{equation}
\label{eq:vev&l_G2}
\chi^{(0)} = \pm \sqrt{\frac{\sqrt{65}-5}{2}}
\quad , \quad
\ell^{2} = \frac{63}{64} (9-\sqrt{65}) \ .
\end{equation}
We note in passing that
\begin{equation}
\label{eq:l_ordering}
\ell^{2}_{\SO{7}} \, (=1) \,>\, \ell^{2}_{\SO{6}} \, (\sim 0.929) \,>\, \ell^{2}_{\GG_{2}} \, (\sim 0.923) \ ,   
\end{equation}
which suggests that a net of generalised RG-flows might exist connecting the corresponding dual QFT$_{3}$s with a generalised conformal symmetry, in analogy with the RG-flows in massive IIA constructed in \cite{Guarino:2016ynd,Bagshaw:2025mcw}.

At this point, one may ask whether it is also possible to uplift the G$_2$-invariant model in (\ref{eq:Lagrangian_G2}) with $m=0$ to an auxiliary theory in $(4+\eta)$ dimensions with $\eta=1/3$. Keeping the same KK Ansatz as in (\ref{eq:aux_metric}), an explicit computation shows that (the brane frame version of) (\ref{eq:Lagrangian_G2}) is obtained from the auxiliary theory
\begin{equation}\label{eq:aux_G2}
\begin{aligned}
S_{\textrm{G}_{2}}^{(\text{aux})} &= \frac{\calR^{2}}{g^{2}} \vol_{\eta}^{-1} \int \dd^{4}x\,  \dd^{\eta} \xi \sqrt{-g_{4+\eta}}  \\
& \left( R - \frac{1}{2} F_{\eta+1} \wedge \star F_{\eta+1}  -\frac{\calR^{2}}{2}  V_{\chi_{\eta}}  \right) \ ,
\end{aligned}
\end{equation}
where 
\begin{equation}
F_{\eta+1} = \dd \chi_{\eta}
\qquad \textrm{with} \qquad
\chi_{\eta} = \chi \, e^{-\frac{2}{5} \hat{\sigma}} \, \vol(\mathbb{R}^{\eta}) \ , 
\end{equation}
and $\hat{\sigma}$ is given in (\ref{hat_sigma_def}). Note that we have promoted the original axion $\chi$ to an $\eta$-form, $\chi_{\eta}$, in the auxiliary space, in the spirit of \cite{Biggs:2023sqw}. The brane-frame scalar potential $V_{\chi_{\eta}}$ in (\ref{eq:aux_G2}) keeps the form of $V_{\chi}$ in (\ref{V_chi}), with the replacement $\chi^{2} \rightarrow |\chi_{\eta}|^{2}$, where $|\chi_{\eta}|^{2} \equiv \frac{1}{\eta!} \, \chi_{i_{1}\ldots i_{\eta}} \, \chi^{i_{1}\ldots i_{\eta}} = \chi^{2}$. The explicit dependence of the potential on $\chi_{\eta}$ breaks gauge invariance, so $\chi_{\eta}$ behaves as a Proca-like field.

Interestingly, the uplift of (\ref{eq:confAdSAxion})-(\ref{eq:vev&l_G2}) to the auxiliary theory in (\ref{eq:aux_G2}) does not lead to a round AdS$_{4+\eta}$ solution with $\eta=1/3$, but rather to a Lifshitz-like geometry, where the auxiliary $\eta$-dimensional torus has anisotropic scaling
\begin{equation}\label{eq:G2Lifshitz}
\dd s^{2}_{4+\eta} = (\ell_{\GG_{2}} r)^{2}\dd x^{2}_{1,2} + \ell^{2}_{\GG_{2}}\frac{\dd r^{2}}{r^{2}} + (\ell_{\GG_{2}} r)^{\frac{2z}{\eta}} \delta_{ij}\dd\xi^{i}\dd\xi^{j} \ ,
\end{equation}
with $z=\frac{\sqrt{65}-7}{4}$ \footnote{Still, when restricted to fields independent of $\xi^{I}$, the scalar Laplacian on the spacetime \eqref{eq:G2Lifshitz} coincides with the AdS$_{4+z}$ scalar Laplacian restricted to fields independent of the $z$ extra coordinates.}. As such, it is not possible to interpret this solution as a conformal fixed point in the dual auxiliary CFT$_{3+\eta}$. However, in standard holography, domain-wall solutions that interpolate between AdS and Lifshitz fixed points are known, \textit{e.g.} \cite{Bertoldi:2011zr} which requires a Proca field, thus, one may wonder whether a similar interpretation is possible here, and what would be the interpretation after dimensionally reducing to the 3D QFT with a generalised conformal symmetry.

Finally, the auxiliary theory in (\ref{eq:aux_G2}), which includes an axion, can be viewed as a first extension of the theory in (\ref{eq:auxAdSGrav}) describing the better-established metric–dilaton–$F_{8-p}$ sector of non-conformal D$p$-brane holography. A further extension to the $\mathbb{Z}_{2}^{3}$-invariant sector of \cite{Guarino:2019snw}, which contains seven axions, is straightforward to derive, and we expect a similar structure to hold for the full field content of ISO(7) maximal supergravity. We hope to return to these questions in future work.

\vspace{2mm}

\noindent\textbf{Acknowledgements}:  We are grateful to Nathan Bagshaw for collaboration on related work, and Carlos Nunez and Guillermo Mera for conversations. The work of A.C. is supported by the Severo Ochoa fellowship PA-23-BP22-019. The work of R.S. is supported by the FWO grants G094523N and G003523N. The work of A.C. and A.G. is supported in part by the grants from the Spanish government MCIU-22-PID2021-123021NB-I00 and MCIU-25-PID2024-161500NB-I00.

\vspace{2mm}

\noindent\textit{Note added:} While preparing this manuscript, we became aware of the related work in \cite{Bobev:2026kld}, which may have some overlap with our results. We have coordinated the submission of our manuscripts with the authors of \cite{Bobev:2026kld}.

\appendix

\section{Parameterising the seven-dilaton sector}
\label{app:scalar_parameterisation}

Starting from the seven dilatons in Appendix~A of \cite{Guarino:2019snw}, $\boldsymbol{\varphi}=(\varphi_{1},\ldots,\varphi_{7})^{T}$, we introduce a new set of scalars, $\boldsymbol{\lambda} = (\sigma,\lambda_{a})^{T}$, with $a=1,\ldots,6$, given by $\boldsymbol{\lambda} = M \boldsymbol{\varphi}$ with
\begin{equation}
M =
\resizebox{0.4\textwidth}{!}{$
\begin{pmatrix}
\frac{1}{\sqrt{7}} &
\frac{1}{\sqrt{7}} &
\frac{1}{\sqrt{7}} &
\frac{1}{\sqrt{7}} &
\frac{1}{\sqrt{7}} &
\frac{1}{\sqrt{7}} &
\frac{1}{\sqrt{7}}
\\[4pt]
0 & \frac{1}{2} & \frac{1}{2} & -\frac{1}{2} & 0 & 0 & -\frac{1}{2}
\\[4pt]
0 & \frac{1}{2\sqrt{3}} & \frac{1}{2\sqrt{3}} & \frac{1}{2\sqrt{3}} &
-\frac{1}{\sqrt{3}} & -\frac{1}{\sqrt{3}} & \frac{1}{2\sqrt{3}}
\\[4pt]
\frac{3}{2\sqrt{6}} & -\frac{1}{\sqrt{6}} & \frac{1}{2\sqrt{6}} &
\frac{1}{2\sqrt{6}} & -\frac{1}{\sqrt{6}} & \frac{1}{2\sqrt{6}} &
-\frac{1}{\sqrt{6}}
\\[4pt]
\frac{3}{2\sqrt{10}} & -\frac{1}{\sqrt{10}} &
\frac{1}{2\sqrt{10}} & -\frac{3}{2\sqrt{10}} &
\frac{1}{\sqrt{10}} & -\frac{3}{2\sqrt{10}} &
\frac{1}{\sqrt{10}}
\\[4pt]
\frac{3}{2\sqrt{15}} & \frac{3}{2\sqrt{15}} &
-\frac{2}{\sqrt{15}} & \frac{1}{\sqrt{15}} &
\frac{1}{\sqrt{15}} & -\frac{3}{2\sqrt{15}} &
-\frac{3}{2\sqrt{15}}
\\[4pt]
\frac{3}{2\sqrt{21}} & \frac{3}{2\sqrt{21}} &
-\frac{2}{\sqrt{21}} & -\frac{2}{\sqrt{21}} &
-\frac{2}{\sqrt{21}} & \frac{3}{2\sqrt{21}} &
\frac{3}{2\sqrt{21}}
\end{pmatrix}
$} \ .
\end{equation}
The $\lambda_{a}$ scalars are those appearing in the r.h.s of (\ref{eq:4D_Lagrangian_dual}), and they have canonically normalised kinetic terms. From the $\lambda_{a}$ scalars, we have also introduced auxiliary scalars $ X_{I} \equiv e^{-\frac{1}{2} (b_{I})^{a} \lambda_{a}}$ using the weights, $(b_{I})^{a}$, of the fundamental representation of $\textrm{SL}(7)$, which are given by
\begin{equation}
    (b_{I})^{a} = 
    \begin{pmatrix}
        2&\frac{2}{\sqrt{3}}&\sqrt{\frac{2}{3}}&\sqrt{\frac{2}{5}}&\frac{2}{\sqrt{15}}&\frac{2}{\sqrt{21}}\\
        -2&\frac{2}{\sqrt{3}}&\sqrt{\frac{2}{3}}&\sqrt{\frac{2}{5}}&\frac{2}{\sqrt{15}}&\frac{2}{\sqrt{21}}\\
        0&-\frac{4}{\sqrt{3}}&\sqrt{\frac{2}{3}}&\sqrt{\frac{2}{5}}&\frac{2}{\sqrt{15}}&\frac{2}{\sqrt{21}}\\
        0&0&-\sqrt{6}&\sqrt{\frac{2}{5}}&\frac{2}{\sqrt{15}}&\frac{2}{\sqrt{21}}\\
        0&0&0&-4 \sqrt{\frac{2}{5}}&\frac{2}{\sqrt{15}}&\frac{2}{\sqrt{21}}\\
        0&0&0&0&-2 \sqrt{\frac{5}{3}}&\frac{2}{\sqrt{21}}\\
        0&0&0&0&0&-4 \sqrt{\frac{3}{7}}
    \end{pmatrix} .
\end{equation}
They satisfy $\prod_{I} X_{I}=1$, and were used, for example, to present the seven-dilaton Lagrangian (\ref{eq:7DilatonLagrangian}).

\section{IIA uplift of the seven-dilaton model}
\label{app:IIA_uplift_7_dilatons}

Here we present the uplift of the seven-dilaton model (\ref{eq:7DilatonLagrangian}) to massive type IIA supergravity in ten dimensions. In string frame, the field content of (\ref{eq:7DilatonLagrangian}) is embedded into massive IIA supergravity as
\begin{align}\label{eq:Uplift7Dilaton}
        \dd s^{2}_{10} &= e^{\frac{2}{\sqrt{7}}\sigma}\Delta^{\frac{1}{2}} \dd s^{2}_{4} + \left(\frac{\sqrt{2}}{g}\right)^{2}e^{\frac{\sigma}{\sqrt{7}}} \Delta^{-\frac{1}{2}} \sum^{7}_{I=1} X_{I}^{-1} (\dd \mu_{I})^{2} , \nonumber \\
        e^{\Phi} &= e^{\frac{5}{2\sqrt{7}}\sigma} \Delta^{-\frac{1}{4}} \quad , \quad F_{0} = \frac{m}{\sqrt{2}} \ , \\
        F_{4} &= -\frac{g}{\sqrt{2}} e^{\frac{\sigma}{\sqrt{7}}} \sum^{7}_{I=1} \left( 2 \, X_{I}^{2} \,\mu_{I}^{2} - \Delta \, X_{I} \right) \text{vol}_4 \nonumber \\
        &-\frac{1}{2}\frac{\sqrt{2}}{g} \sum^{7}_{I=1} X_{I}^{-1} \star_{4}\dd X_{I} \wedge \dd( \mu^{2}_{I} ) \ , \nonumber
\end{align}
where $\mu_{I}$, with $I=1,\ldots,7$, are embedding coordinates on the six-sphere satisfying $\sum_{I}\mu_{I}^{2}=1$,  $\Delta = \sum_{I} X_{I} \mu_{I}^{2}$ and $\text{vol}_4$ is the volume form of the four-dimensional supergravity metric in (\ref{eq:7DilatonLagrangian}). In the massless case, $m=0$, this uplift reduces to the one constructed in \cite{Cvetic:2000zu}.

\subsection{Uplift of solutions in Section~\ref{sec:flatDW}}

The massless type IIA uplift of the generalised conformal solutions in Section~\ref{sec:flatDW} takes the form
\begin{align}
\label{eq:uplift_generalised_conformal}
    \dd s^2_{10} & = \left( \frac{\sqrt{2}}{g}\right)^2 e^{\frac{\sigma}{\sqrt{7}}} \bigg[ \Delta^{\frac{1}{2}} \, \calR^2 \, \ell^2 \dd s^2_{\text{AdS}_4} \nonumber \\
    & + Y^{\frac{1}{6}} \Delta^{-\frac{1}{2}} \bigg( Y^{-1} \Delta \, \dd \alpha^2 + \sin^2 \alpha \, \dd s^2_{\text{S}^5} \bigg)\bigg] \ , \nonumber \\
    e^{\Phi} &= e^{ \frac{5 \sigma}{2 \sqrt{7}}} \Delta^{- \frac{1}{4}} \quad,\quad  F_0 = 0 \ ,  \\
    F_4 & = \frac{g}{\sqrt{2}} e^{\frac{\sigma}{\sqrt{7}}}  \bigg(\Delta \left( Y + 6 \, Y^{-\frac{1}{6}} \right) \nonumber \\
    & - 2 \, Y^{-\frac{1}{3}} \left(\sin^2 \alpha + Y^{\frac{7}{3}}  \cos^2 \alpha \right) \bigg) \text{vol}_4 \ , \nonumber
\end{align}
where $\Delta =Y^{-\frac{1}{6}} \left(\sin^2 \alpha + Y^{\frac{7}{6}}  \cos^2 \alpha \right)$. While the round SO(7)-symmetric solution has $Y=1$, the deformed SO(6)-symmetric solutions have $Y=2^{\frac{12}{7}}$.

\subsection{Uplift of the \texorpdfstring{$\textrm{SO}(n) \times \textrm{SO}(7-n)$}{SO(n) x SO(7-n)} invariant sector}
\label{app:Coulomb_branch_SO(n)xSO(7-n)_uplift}

The massless type IIA uplift of the field content in Section~\ref{sec:Coulomb_branch_SO(n)xSO(7-n)} reads
    \begin{align}
        \dd s_{10}^{2} & =  e^{\frac{2}{\sqrt{7}}\sigma}\Delta_{\varphi}^{\frac{1}{2}} \, \dd s_{4}^{2} 
        + \frac{2}{g^{2}}e^{\frac{\sigma}{\sqrt{7}}+\frac{1}{c_{n}}((7-n)-n)\varphi} \Delta_{\varphi}^{\frac{1}{2}}  
        \dd\alpha^{2} \nonumber \\
        &\phantom{=} +\frac{2}{g^{2}}e^{\frac{\sigma}{\sqrt{7}}} \Delta_{\varphi}^{-\frac{1}{2}}\left[ 
        e^{\frac{7-n}{c_{n}}\varphi}\cos^{2}\alpha \, \dd\Omega^{2}_{n-1} \right. \nonumber \\
        &\left. \,\,\,\, + \,e^{-\frac{n}{c_{n}}\varphi} \sin^{2}\alpha \, \dd\Omega^{2}_{7-n-1} \right] \ , \\
        e^{\Phi} &= e^{\frac{5}{2\sqrt{7}}\sigma} \Delta_{\varphi}^{-\frac{1}{4}} \quad , \quad F_{0} = 0 \ , \nonumber \\
        F_{4} &= \frac{g}{\sqrt{2}} e^{\frac{\sigma}{\sqrt{7}}}f_{0}\,\text{vol}_{4} + \frac{7\sqrt{2}}{2c_{n}\, g}\sin(2\alpha)\dd \alpha \wedge \star_{4} \dd \varphi \ , \nonumber
    \end{align}
where $\dd\Omega^{2}_{n}$ is the line element of a unit radius $n$-sphere (in this notation $\dd\Omega^{2}_{0}=0$), and
    \begin{align}
        &f_{0} = 2\left(e^{-\frac{2(7-n)}{c_{n}}\varphi}\cos^{2}\alpha + e^{\frac{2n}{c_{n}}\varphi}\sin^{2}\alpha \right) \nonumber \\
        & \quad - \Delta_{\varphi} \left( n\, e^{-\frac{7-n}{c_{n}}\varphi} + (7-n)\, e^{-\frac{n}{c_{n}}\varphi} \right) \ ,\\
        &\Delta_{\varphi} = e^{-\frac{7-n}{c_{n}}\varphi} \cos^{2}\alpha +e^{\frac{n}{c_{n}}\varphi} \sin^{2}\alpha \ . \nonumber
    \end{align}
Note that the structure of the uplift is invariant under a simultaneous transformations
\begin{equation}
n \leftrightarrow 7-n
\quad,\quad
\varphi \rightarrow - \varphi
\quad,\quad
\alpha \rightarrow  \frac{\pi}{2}-\alpha \ .
\end{equation}

\section{Curved domain-walls in massive IIA}\label{app:curvedDW}
\label{app:curved_DW_mIIA}

In the massive case $m \neq 0$, there exists a class of \textit{curved-sliced} domain-walls with constant $X_{I}=X^{(0)}_{I}$ and a running $\sigma$ of the form
\begin{equation}
\label{eq:AdS3DomainWallMassive}
\dd s^{2}_{4} = \calR^2 \left(\frac{\sqrt{2}}{g}\right)^{2} e^{- \frac{\sigma}{\sqrt{7}}}  \dd s^{2}_{\textrm{AdS}_{4}} 
\,\,\, , \,\,\, 
e^{-\frac{\sigma}{\sqrt{7}}} = \left( \calR\frac{\sqrt{2}}{g} \, \ell  \, r \right)^{z} ,
\end{equation}
with $z=1/3$, and
\begin{equation}
\label{eq:AdS4_metricMassive}
\dd s^{2}_{\textrm{AdS}_{4}}  = \ell^{2} \left( \left( r^2 + \frac{9 m^2}{8}\right) \dd s^2_{\textrm{AdS}_{3}} + \frac{\dd r^2}{\left( r^2 + \frac{9 m^2}{8}\right)} \right) ,
\end{equation}    
where, in these conventions, AdS$_3$ is parameterised such that $R_{\mu \nu}^{\text{AdS}_3} = - 2 \left(\frac{9 m^2}{8}\right) g_{\mu \nu}^{\text{AdS}_3}$. As in Section~\ref{sec:flatDW}, we have $\calR = 2/3$ and an AdS$_{4}$ radius given by $\ell^{2} = - 35/V^{(0)}_{X}$. We note that taking $m \to 0$, the curved-sliced domain-walls smoothly reduce to their flat-sliced counterparts. Consistently with this limit, the SO(7)-symmetric domain-wall constructed from (\ref{eq:XSO7vacuum}) is $1/2$-BPS supersymmetric, whereas the SO(6)-symmetric ones constructed from (\ref{eq:XSO6vacuum}) are non-supersymmetric. The uplift of (\ref{eq:AdS3DomainWallMassive})-(\ref{eq:AdS4_metricMassive}) is still given by (\ref{eq:uplift_generalised_conformal}) but with $F_{0} = \frac{m}{\sqrt{2}}$.

\section{BPS-equations}
\label{app:BPS_equations}

For $m=0$, we derive the BPS-equations associated with a flat-sliced domain-wall of the form (\ref{eq:DW-ansatz_main_text}), and with all scalars depending only on the radial coordinate $\rho$. The Ansatz (\ref{eq:DW-ansatz_main_text}) accommodates the metric in \eqref{eq:D2braneConstantLambda}-\eqref{eq:AdS4_metric} and also the one in (\ref{eq:Ansatz4DCoulomb}) after suitable redefinitions of the radial coordinate.

\subsection{The seven-dilaton model}
\label{app:BPS_equations_7_dilaton}

In the seven-dilaton model (\ref{eq:7DilatonLagrangian}), the scalar potential follows from a factorised superpotential
\begin{equation}
W = \frac{g}{4\sqrt{2}} \, e^{\frac{\sigma}{2\sqrt{7}}} \, W_{\lambda}
\quad \textrm{ with } \quad
W_{\lambda} = \sum_{I }e^{-\frac{1}{2} (b_{I})^{a} \lambda_{a}} \ .
\end{equation}
Assuming scalars of the form $\sigma = \sigma(\rho)$ and $\lambda_{a} = \lambda_{a}(\rho)$, a set of first-order BPS-equations follows of the form
\begin{equation}
\label{eq:4DBPSlambdasigma}
B' = \frac{3\calR}{14} W_{\lambda}
\,\,\,,\,\,\, 
\sigma' = - \frac{\calR}{2\sqrt{7}} W_{\lambda}
\,\,\,,\,\,\,
\lambda'_{a} = - \calR \frac{\partial W_{\lambda}}{\partial \lambda_{a}} \ ,
\end{equation}
where the prime indicates a derivative with respect to the radial coordinate $\rho$. First, note that since $W_{\lambda}>0$, so that there are no supersymmetric solutions with constant $B$ and $\sigma$, which is expected. On the other hand, the system allows for supersymmetric solutions of constant $\lambda_{a}$ if $\frac{\partial W_{\lambda}}{\partial \lambda_{a}}=0$, which is only satisfied for $\lambda_{a}=0$, and corresponds to the SO(7)-symmetric D2-brane solution in (\ref{eq:XSO7vacuum}). Finally, for this family of domain-wall solutions, the BPS-equations lead to the following relation between the scalar $\sigma$ and the metric function $B$, 
\begin{equation}
\label{eq:Bandsigma}
\sigma = -\frac{\sqrt{7}}{3} B + \sigma_{0} \ ,
\end{equation}
with $\sigma_{0}$ being a real integration constant.

\subsection{The \texorpdfstring{G$_{2}$}{G2}-invariant sector}
\label{app:BPS_equations_G2-sector}

Starting from the G$_{2}$-invariant model (\ref{eq:Lagrangian_G2}) and assuming scalars of the form $\sigma  = \sigma(\rho)$ and $\chi =\chi(\rho)$, the BPS-equations read 
\begin{equation}
B'(\rho)  = \frac{3\calR}{2\sqrt{7}} \sqrt{7+\chi^{2}} \ ,
\end{equation}
and
\begin{equation}
\sigma'(\rho)  = -\frac{\calR}{2}\frac{7+8\chi^{2} +\chi^{4}}{\sqrt{7+\chi^{2}}}
\,\,,\,\,
\chi'(\rho)  = -\frac{\calR}{2\sqrt{7}} \chi \left(7+\chi^{2}\right)^{\frac{3}{2}} \ .
\end{equation}
From this, it becomes clear that there are no supersymmetric solutions with a constant axion $\chi^{(0)} \neq 0$.

\bibliographystyle{apsrev4-1}
\bibliography{references}

\providecommand{\href}[2]{#2}\begingroup\raggedright\begin{thebibliography}{10}

\bibitem{Hull:1988jw}
C.~M. Hull and N.~P. Warner, ``{Noncompact Gaugings From Higher Dimensions},'' \href{http://dx.doi.org/10.1088/0264-9381/5/12/005}{{\em Class. Quant. Grav.} {\bfseries 5} (1988) 1517}.

\bibitem{Hull:1984yy}
C.~M. Hull, ``{A New Gauging of $N=8$ Supergravity},'' \href{http://dx.doi.org/10.1103/PhysRevD.30.760}{{\em Phys. Rev. D} {\bfseries 30} (1984) 760}.

\bibitem{Romans:1985tz}
L.~J. Romans, ``{Massive N=2a Supergravity in Ten-Dimensions},'' \href{http://dx.doi.org/10.1016/0370-2693(86)90375-8}{{\em Phys. Lett. B} {\bfseries 169} (1986) 374}.

\bibitem{Guarino:2015vca}
A.~Guarino and O.~Varela, ``{Consistent $ \mathcal{N}=8 $ truncation of massive IIA on S$^{6}$},'' \href{http://dx.doi.org/10.1007/JHEP12(2015)020}{{\em JHEP} {\bfseries 12} (2015) 020}, \href{http://arxiv.org/abs/1509.02526}{{\ttfamily arXiv:1509.02526 [hep-th]}}.

\bibitem{Guarino:2015qaa}
A.~Guarino and O.~Varela, ``{Dyonic ISO(7) supergravity and the duality hierarchy},'' \href{http://dx.doi.org/10.1007/JHEP02(2016)079}{{\em JHEP} {\bfseries 02} (2016) 079}, \href{http://arxiv.org/abs/1508.04432}{{\ttfamily arXiv:1508.04432 [hep-th]}}.

\bibitem{Kanitscheider:2008kd}
I.~Kanitscheider, K.~Skenderis, and M.~Taylor, ``{Precision holography for non-conformal branes},'' \href{http://dx.doi.org/10.1088/1126-6708/2008/09/094}{{\em JHEP} {\bfseries 09} (2008) 094}, \href{http://arxiv.org/abs/0807.3324}{{\ttfamily arXiv:0807.3324 [hep-th]}}.

\bibitem{Jevicki:1998yr}
A.~Jevicki and T.~Yoneya, ``{Space-time uncertainty principle and conformal symmetry in D particle dynamics},'' \href{http://dx.doi.org/10.1016/S0550-3213(98)00578-1}{{\em Nucl. Phys. B} {\bfseries 535} (1998) 335--348}, \href{http://arxiv.org/abs/hep-th/9805069}{{\ttfamily arXiv:hep-th/9805069}}.

\bibitem{Jevicki:1998ub}
A.~Jevicki, Y.~Kazama, and T.~Yoneya, ``{Generalized conformal symmetry in D-brane matrix models},'' \href{http://dx.doi.org/10.1103/PhysRevD.59.066001}{{\em Phys. Rev. D} {\bfseries 59} (1999) 066001}, \href{http://arxiv.org/abs/hep-th/9810146}{{\ttfamily arXiv:hep-th/9810146}}.

\bibitem{Guarino:2015jca}
A.~Guarino, D.~L. Jafferis, and O.~Varela, ``{String Theory Origin of Dyonic N=8 Supergravity and Its Chern-Simons Duals},'' \href{http://dx.doi.org/10.1103/PhysRevLett.115.091601}{{\em Phys. Rev. Lett.} {\bfseries 115} no.~9, (2015) 091601}, \href{http://arxiv.org/abs/1504.08009}{{\ttfamily arXiv:1504.08009 [hep-th]}}.

\bibitem{Note1}
Still we have verified that the results here extend straightforwardly to the STU-model of \cite {Guarino:2017pkw} with vanishing axions.

\bibitem{Biggs:2023sqw}
A.~Biggs and J.~Maldacena, ``{Scaling similarities and quasinormal modes of D0 black hole solutions},'' \href{http://dx.doi.org/10.1007/JHEP11(2023)155}{{\em JHEP} {\bfseries 11} (2023) 155}, \href{http://arxiv.org/abs/2303.09974}{{\ttfamily arXiv:2303.09974 [hep-th]}}.

\bibitem{Bobev:2025idz}
N.~Bobev, G.~Mera~{\'A}lvarez, and H.~Paul, ``{Correlation functions for non-conformal Dp-brane holography},'' \href{http://dx.doi.org/10.1007/JHEP07(2025)137}{{\em JHEP} {\bfseries 07} (2025) 137}, \href{http://arxiv.org/abs/2503.18770}{{\ttfamily arXiv:2503.18770 [hep-th]}}.

\bibitem{BBGM}
N.~Bobev, P.~Bomans, F.~F. Gautason, and G.~Mera~Álvarez, ``{Holographic Correlators for Non-Conformal Maximally Supersymmetric Yang-Mills},'' {\em {to appear}} .

\bibitem{Gubser:2000nd}
S.~S. Gubser, ``{Curvature singularities: The Good, the bad, and the naked},'' \href{http://dx.doi.org/10.4310/ATMP.2000.v4.n3.a6}{{\em Adv. Theor. Math. Phys.} {\bfseries 4} (2000) 679--745}, \href{http://arxiv.org/abs/hep-th/0002160}{{\ttfamily arXiv:hep-th/0002160}}.

\bibitem{Guarino:2019snw}
A.~Guarino, J.~Tarrio, and O.~Varela, ``{Flowing to $\mathcal{N}=3$ Chern-Simons-matter theory},'' \href{http://dx.doi.org/10.1007/JHEP03(2020)100}{{\em JHEP} {\bfseries 03} (2020) 100}, \href{http://arxiv.org/abs/1910.06866}{{\ttfamily arXiv:1910.06866 [hep-th]}}.

\bibitem{Cvetic:2000zu}
M.~Cvetic, H.~Lu, and C.~N. Pope, ``{Consistent sphere reductions and universality of the Coulomb branch in the domain wall / QFT correspondence},'' \href{http://dx.doi.org/10.1016/S0550-3213(00)00462-4}{{\em Nucl. Phys. B} {\bfseries 590} (2000) 213--232}, \href{http://arxiv.org/abs/hep-th/0004201}{{\ttfamily arXiv:hep-th/0004201}}.

\bibitem{Guarino:2019jef}
A.~Guarino, J.~Tarr{\'\i}o, and O.~Varela, ``{Halving ISO(7) supergravity},'' \href{http://dx.doi.org/10.1007/JHEP11(2019)143}{{\em JHEP} {\bfseries 11} (2019) 143}, \href{http://arxiv.org/abs/1907.11681}{{\ttfamily arXiv:1907.11681 [hep-th]}}.

\bibitem{Bobev:2020qev}
N.~Bobev, T.~Fischbacher, F.~F. Gautason, and K.~Pilch, ``{New AdS$_4$ Vacua in Dyonic ISO(7) Gauged Supergravity},'' \href{http://arxiv.org/abs/2011.08542}{{\ttfamily arXiv:2011.08542 [hep-th]}}.

\bibitem{Note2}
Still, massless IIA solutions with a running dilaton and an AdS$_{4}$ factor in string frame have been discussed in \cite {Gaillard:2010gy}.

\bibitem{Dibitetto:2018ftj}
G.~Dibitetto, G.~Lo~Monaco, A.~Passias, N.~Petri, and A.~Tomasiello, ``{AdS$_3$ Solutions with Exceptional Supersymmetry},'' \href{http://dx.doi.org/10.1002/prop.201800060}{{\em Fortsch. Phys.} {\bfseries 66} no.~10, (2018) 1800060}, \href{http://arxiv.org/abs/1807.06602}{{\ttfamily arXiv:1807.06602 [hep-th]}}.

\bibitem{Legramandi:2020txf}
A.~Legramandi, G.~Lo~Monaco, and N.~T. Macpherson, ``{All $\mathcal{N}=(8,0)$ AdS$_3$ solutions in 10 and 11 dimensions},'' \href{http://dx.doi.org/10.1007/JHEP05(2021)263}{{\em JHEP} {\bfseries 05} (2021) 263}, \href{http://arxiv.org/abs/2012.10507}{{\ttfamily arXiv:2012.10507 [hep-th]}}.

\bibitem{Boonstra:1998mp}
H.~J. Boonstra, K.~Skenderis, and P.~K. Townsend, ``{The domain wall / QFT correspondence},'' \href{http://dx.doi.org/10.1088/1126-6708/1999/01/003}{{\em JHEP} {\bfseries 01} (1999) 003}, \href{http://arxiv.org/abs/hep-th/9807137}{{\ttfamily arXiv:hep-th/9807137}}.

\bibitem{Breitenlohner:1982jf}
P.~Breitenlohner and D.~Z. Freedman, ``{Stability in Gauged Extended Supergravity},'' \href{http://dx.doi.org/10.1016/0003-4916(82)90116-6}{{\em Annals Phys.} {\bfseries 144} (1982) 249}.

\bibitem{Note3}
Using (\ref {eq:XtoSectors}), the solutions (\ref {eq:Ansatz4DCoulomb})-(\ref {eq:Solution4Dv1}) solve the BPS-equations in Appendix~\ref {app:BPS_equations_7_dilaton} with a factorised real superpotential $W = \protect \frac {g}{4 \protect \sqrt {2}} e^{\protect \frac {\sigma }{2 \protect \sqrt {7}}} \protect \, W_{\varphi }$ with $W_{\varphi } = n \protect \, e^{- \protect \frac {7-n}{c_{n}} \protect \, \varphi } + (7-n) \protect \, e^{\protect \frac {n}{c_{n}} \protect \, \varphi }$. We note that $c_{n}$ is symmetric under the exchange $n \leftrightarrow (7-n)$, so the four-dimensional theory is invariant under the simultaneous transformations $\varphi \rightarrow - \varphi $ and $n \leftrightarrow (7-n)$. However, the solutions (\ref {eq:Ansatz4DCoulomb})-(\ref {eq:Solution4Dv1}) are not.

\bibitem{Freedman:1999gk}
D.~Z. Freedman, S.~S. Gubser, K.~Pilch, and N.~P. Warner, ``{Continuous distributions of D3-branes and gauged supergravity},'' \href{http://dx.doi.org/10.1088/1126-6708/2000/07/038}{{\em JHEP} {\bfseries 07} (2000) 038}, \href{http://arxiv.org/abs/hep-th/9906194}{{\ttfamily arXiv:hep-th/9906194}}.

\bibitem{Maldacena:2000mw}
J.~M. Maldacena and C.~Nunez, ``{Supergravity description of field theories on curved manifolds and a no go theorem},'' \href{http://dx.doi.org/10.1142/S0217751X01003937}{{\em Int. J. Mod. Phys. A} {\bfseries 16} (2001) 822--855}, \href{http://arxiv.org/abs/hep-th/0007018}{{\ttfamily arXiv:hep-th/0007018}}.

\bibitem{Guarino:2016ynd}
A.~Guarino, J.~Tarrio, and O.~Varela, ``{Romans-mass-driven flows on the D2-brane},'' \href{http://dx.doi.org/10.1007/JHEP08(2016)168}{{\em JHEP} {\bfseries 08} (2016) 168}, \href{http://arxiv.org/abs/1605.09254}{{\ttfamily arXiv:1605.09254 [hep-th]}}.

\bibitem{Bagshaw:2025mcw}
N.~Bagshaw and G.~Dibitetto, ``{Non-SUSY DW{\textquoteright}s in ISO(7) gauged supergravity},'' \href{http://dx.doi.org/10.1007/JHEP07(2026)052}{{\em JHEP} {\bfseries 07} (2026) 052}, \href{http://arxiv.org/abs/2512.24697}{{\ttfamily arXiv:2512.24697 [hep-th]}}.

\bibitem{Note4}
Still, when restricted to fields independent of $\xi ^{I}$, the scalar Laplacian on the spacetime \protect \eqref {eq:G2Lifshitz} coincides with the AdS$_{4+z}$ scalar Laplacian restricted to fields independent of the $z$ extra coordinates.

\bibitem{Bertoldi:2011zr}
G.~Bertoldi, B.~A. Burrington, A.~W. Peet, and I.~G. Zadeh, ``{Lifshitz-like black brane thermodynamics in higher dimensions},'' \href{http://dx.doi.org/10.1103/PhysRevD.83.126006}{{\em Phys. Rev. D} {\bfseries 83} (2011) 126006}, \href{http://arxiv.org/abs/1101.1980}{{\ttfamily arXiv:1101.1980 [hep-th]}}.

\bibitem{Bobev:2026kld}
N.~Bobev, P.~Bomans, and F.~F. Gautason, ``{Scale-Covariant Holography},'' \href{http://arxiv.org/abs/2609.38302}{{\ttfamily arXiv:2609.38302 [hep-th]}}.

\bibitem{Guarino:2017pkw}
A.~Guarino, ``{BPS black hole horizons from massive IIA},'' \href{http://dx.doi.org/10.1007/JHEP08(2017)100}{{\em JHEP} {\bfseries 08} (2017) 100}, \href{http://arxiv.org/abs/1706.01823}{{\ttfamily arXiv:1706.01823 [hep-th]}}.

\bibitem{Gaillard:2010gy}
J.~Gaillard and D.~Martelli, ``{Fivebranes and resolved deformed $G_2$ manifolds},'' \href{http://dx.doi.org/10.1007/JHEP05(2011)109}{{\em JHEP} {\bfseries 05} (2011) 109}, \href{http://arxiv.org/abs/1008.0640}{{\ttfamily arXiv:1008.0640 [hep-th]}}.

\end{thebibliography}\endgroup

\end{document}